# Measuring the characteristics of electroosmotic flow in a polyelectrolyte grafted nanopore by molecular theory approach

Milad Reshadi[1], Mohammad Hassan Saidi[*,1]

[1] Center of Excellence in Energy Conversion (CEEC), School of Mechanical Engineering,
Sharif University of Technology, Tehran 11155-9567, Iran

**Abstract**

In this paper, we present a molecular theory analysis of ions and potential distribution, degree of ionization of polyelectrolyte (PE) brushes, velocity profile, volumetric flow rate, ionic selectivity, ionic conduction and advection by electroosmotic flow in poly-acid (PA)/poly-base (PB) grafted nanopores. The generated conformations by the Rotational Isomeric State model are used in performing the minimization of the free energy functional of the system including the effects of the Born energy arising from the variation of permittivity, pH of the electrolyte, grafting density of weak PE brushes, ion partitioning and ionic size. Then, the velocity field is obtained in the process of solving the Navier–Stokes–Brinkman equation by considering the interfacial fluid/wall slippage. Also, the accuracy of the numerical solutions is examined by comparing the present results of ionic conductivity with the existing experimental data for a nanopore grafted with 4PVP brushes used as a synthetic proton-gated ion channel. The application of the present methodology enables us to describe the electrohydrodynamic characteristics of electroosmotic flow in PE grafted nanopores in terms of different factors including the pH of the electrolyte, bulk salt concentration and the grafting density of PE brushes. We show that the dependency of the quantities of interest are essentially rely on the type of the polymer chains. For instance, increasing the pH of the electrolyte results in an increase/decrease of the degree of charged sites in the PA/PB brushes, and there exists a minimum/maximum point in the variation of magnitude of the ion selectivity of PA/PB grafted nanopore with the pH of the electrolyte. However, for both types of PA and PB grafted nanopores, ionic conduction and advection are approximately the ascending function of the bulk salt concentration and low range grafting density of the PE layer.

**Keywords:** molecular theory, polyelectrolyte, ionic selectivity, ionic conduction, ionic advection

[*] Corresponding author. Tel.: +98 21 66165522
*E-mail addresses*: reshadi@mech.sharif.edu (M. Reshadi), saman@sharif.edu (M.H. Saidi)

# 1. Introduction

More than two decades of study in the field of fluid transport in soft nanopores have revealed the immense potential of surface physiochemical modification by polycationic and polyanionic chains in improving the performance and efficiency of nanofluidic devices. The channels in these devices comprise a layer of charged biomolecules called polyelectrolyte layer (PEL) that is localized at the solid-liquid interface, and depending on the grafting density, this layer takes the form of a brush like configuration [1]. The binding between the solid surface and the polymer chains is attained by various techniques like layer by layer polyelectrolyte (PE) deposition method [2, 3], self-assembly of thiolated biomolecules onto the solid surface [4], *etc*. The utilization of soft surface coatings in lab-on-a-chip equipment is of great interest in various engineering and biological applications like diodes and transistors [5, 6], nanoionic valves [7], nanofluidic energy converters [8], cell membranes [9], electroosmotic flow (EOF) suppressors [10], targeted drug delivery processes [11], biosensors [12], current rectifiers [13], *etc*. Also, by taking advantage of polymer properties such as responding to external stimuli [14], regulating the wettability of surfaces [15], and mimicking the structure and dynamics of soft biological organisms [1, 16-19], a novel pathway is provided for creating the nanofluidic devices involving single molecule detection, separation, and other chemical and biological assays.

The investigation of charged soft interface in lab-on-a-chip devices is of ultimate importance in a variety of micro-nanofluidic applications specially those involving electrokinetic (EK) phenomena such as electrophoretic and diffusiophoretic mobility of soft particles [20], diffusioosmosis and EOF in soft nanochannels [10, 21-23], *etc*. Electrokinetic actuation of flows through micro-nanofluidic systems is based on the interaction between the electrical double layer (EDL) and a tangential electric field near a charged ionizable surface[24]. Such tangential electric field can be generated via the external electrodes, axial separation of electrolyte ions due to difference in their diffusion coefficient in the presence of ionic concentration gradient, or may be induced by the *streaming potential (SP)* [25]. The latter potential difference, which is generated due to downstream pressure-driven flow (PDF) of the mobile counterions within the diffuse part of the EDL, creates the basis of two important aspects of EK phenomena, namely, *electroviscous effect* and *electro-chemo-mechanical energy conversion (ECMEC)*. Various experimental [26-28], analytical [21, 22, 29-31] and numerical [8, 10, 32-36] studies have been conducted in recent years to quantify each of aforementioned features in soft nanofluidic devices. The results of these studies have demonstrated that the SP and ECMEC efficiency can be modified several folds by grafting PEL inside the nanofluidic channels.

Controlling the density of PE coating layer to achieve the desired fluidic response is revealed to be useful technique for a variety of soft micro-nanofluidic applications. Determining the properties of non-biological charged soft films [37], interrogating bacterial adhesion to surfaces [38] and colloid stabilization by grafting polymers [39] are of few examples of such PE modification of solid-liquid interface. In general, depending on the density of polymer chains anchored to the surface, the effective permittivity of PEL may be different from that of bulk electrolyte solution [40, 41]. Although the assumption of equal permittivities of PEL-electrolyte phases for sparsely grafted polymer coatings is widely used in most previous theoretical studies [30, 35, 42, 43], the variation of permittivity inside and outside the densely grafted PELs should be taken into account to precisely model the electrohydrodynamics of liquid and ion transport over such soft interfaces [44]. For the sake of characterization and obtaining the distribution of electric potential inside the two interacting PEL-electrolyte phases with different dielectric constants, one may come to the application of the theory of Born electrostatic solvation energy [45] representing a phenomenon called *ion partitioning effect (IPE)*.

Over the last years, some researchers have attempted to study the IPE on electrostatic characteristics of soft membrane systems. The notable investigations on this topic include the identification of effect of solution-membrane permittivity difference on the ion concentration distribution [44, 46], exploring IPE on the motion of particles covered by a dense ion-permeable membrane [40], interrogating the mechanisms of reverse osmosis and nanofiltration in the presence of Born dielectric effect [47-50], *etc*. Among the previous studies, Poddar *et al*. [36] investigated IPE on EK flow characteristics through parallel plate soft nanochannels. Their numerical analysis was conducted in the limiting case of point like electrolyte ions. Moreover, the complex interactions between IP and ionic size, along with their consequent effect on the SP and ECMEC efficiency in the presence of thick overlapping EDL has been explored in our recent work [44].

The influence of various electrostatic phenomena such on hydrodynamics of EK transport has been well documented for rigid nanochannels [51-54]. However, no study of molecular theory approach has been reported to examine the coupled effects of the conformation of the end tethered polymer chains, pH of the electrolyte on charge density of the weak (responsive to pH and salinity of the acidic/basic environment) PE brushes, fluid-wall slippage, IP and ionic size effects on the EK flow characteristics in PE grafted nanopores. In order to accurately evaluate these properties, we generalize our previous analysis of EK flows in soft nanofluidics [44] to include the combined aforementioned effects via molecular theory study. In this approach, the Rotational Isomeric State (RIS) model [55] is employed for the generation of the structure of the brushes containing the poly-acid (PA) groups like polyacrylic acid, or poly-base (PB) monomers such as amines. In contrast to previous works dealing with slit nanochannel [10, 30, 33, 35, 36, 56], the present study focuses on soft circular nanopores which has more resemblance to delivery processes in biological environments and other in-vivo/in-vitro applications such as protein transport through nanoporous membrane [57], DNA and RNA translocation across nuclear pores, drug delivery [58, 59], switchable biological pores [18], *etc*. Then, the effect of IP within a wide range of EK parameters is studied by considering the slip condition of velocity on the surface of the nanopore.

The rest of the paper is organized as follows: in the next section, the flow geometry is elucidated and the physical specification of the problem, the assumptions and the pertinent governing equations are expressed. The description of the molecular theory approach and the numerical aspects of the solution of electric potential and velocity fields are discussed in Sects. 3 to 5. Also, the derivation of the ionic conductance, ionic selectivity of the nanopore, the average degree of PEL charge, and (4) the volumetric flow rate are presented in Sect. 6. After discussing the results in Sect. 7, we end by summarizing our key scientific findings in Sect. 8.

## 2. Problem modeling and formulation

Figure 1 represents the schematic diagram of a soft nanopore of radius $R$ which connects two large reservoirs. We employ the cylindrical coordinate system $(r, z)$ where $r$ denotes the radial distance from the center of the nanopore and $z$ is the axial coordinate. The origin of the coordinate system is located at the center of the nanopore. The flow is considered to be axisymmetric, steady-state, fully developed, incompressible, and laminar. Under such conditions with Reynolds number being much less than unity, the nanopore length ($L$) is taken to be one-three orders of magnitude larger than the nanopore radius to allow the assumption of hydrodynamically developed flow condition within the nearly entire length of the nanopore. The wall of the nanopore is coated by weak polymer brushes which is irreversibly anchored to the inner surface. The PEL contains the distributed positive/negative charges which stem from the ionization reaction of PE molecules and can be regulated by pH degree of the electrolyte solution. This charged soft layer acts as an ion selective semi-permeable membrane [34], such that the ions of electrolyte solution can be present inside and outside the PEL, while the positive/negative charges of this layer remain fixed within it. To generalize the problem, the permittivity of the PEL is taken to be $r$-position dependent and is taken to be different from that of bulk electrolyte.

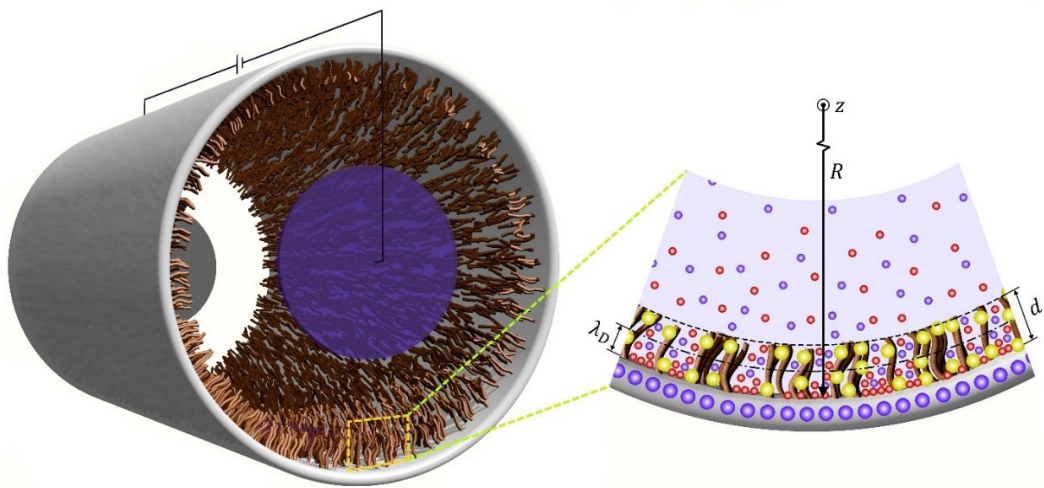

**Fig. 1:** Schematic representation of EOF in the PE-grafted nanopore of radius $R$; the fixed PEL-ions, the charges of the solid surface, the positive and negative ions of the electrolyte solution are illustrated by yellow, large blue, small red and small blue spheres, respectively.

For negatively (or positively) charged PE brushes having ionizable acidic (or basic) groups undergoing the dissociation (or association) reaction $HA \leftrightarrows A^- + H^+$ (or $B + H_2O \leftrightarrows BH^+ + OH^-$), the PEL-liquid electrostatic interaction gives rise to the formation of EDL of thickness $\lambda_D$ equal to the Debye length. In the presence of an applied tangential electric field, the mobile counterions in the diffuse part of the EDL are stimulated to move, and consequently, they exert a viscous drag force to the surrounding liquid resulting in the generation of EK flow in the axial $z$ direction [60]. By employing such flow model, we carry out the analysis of EOF in soft nanopore shown in Fig. 1. The flow under consideration is governed by the set of the continuity and the Navier–Stokes–Brinkman (NSB) equations for fully-developed, unidirectional flow of incompressible ionized fluid with constant physical properties as:

$$\nabla \cdot \boldsymbol{u} = 0 \quad (1)$$

$$\eta \nabla^2 \boldsymbol{u} + \boldsymbol{F_D} + e(\mathbb{z}[\rho_+(r) - \rho_-(r)] + \rho_{H^+}(r) - \rho_{OH^-}(r))\boldsymbol{E} = \boldsymbol{0} \quad (2)$$

where $\boldsymbol{u} = [0,0,u(r)]$ and $\boldsymbol{E} = [0,0,E_z]$ are the velocity and electric field vectors, respectively, $\rho_i(r)$ is the number density of mobile ionic species, $\eta$ is the dynamic viscosity of the liquid, $e$ is the elementary electronic charge and $\mathbb{z}$ is the valence of the electrolyte ions. Due to the presence of PE brushes in the wall adherent regions, there is a hydrodynamic resistance force modeled as $\boldsymbol{F_D} = [0,0,-\mu_c u]$ where $\mu_c = \eta(\langle\phi_p(r)\rangle/a)^2$ is the drag coefficient associated with the resistance to the fluid movement induced by the PEL. In this expression, parameter $a$ is the effective monomer size [61], and $\langle\phi_p(r)\rangle$ is the conformational average of the monomer volume fraction. It is noted that $\langle\phi_p(r)\rangle$ can be obtained from the free energy minimization introduced in the next section.

## 3. Molecular theory approach considering the variation of permittivity

The solution of momentum equation (2) requires the determination of mobile ions distributions, $\rho_i(r)$, which may be obtained by minimizing the free energy functional of the system ($\mathcal{F}$) subject to thermodynamic equilibrium conditions [62]. For a symmetric $\mathbb{z} : \mathbb{z}$ electrolyte, the integral description of $\mathcal{F}$ per unit area of the wall of the PA grafted nanopore per unit thermal energy can be presented as:

$$\frac{\beta \mathcal{F}}{A_0} = \sigma_p \sum_\alpha P(\alpha) \ln P(\alpha) + \frac{1}{R}\Biggl\{\int r dr \rho_w(r)\{\ln \rho_w(r) v_w - 1\} + \sum_{i=\{+,-\}} \int r dr \rho_i(r)(\ln \rho_i(r) v_w - 1) \quad (3)$$

$$+ \sum_{i=\{H^+,OH^-\}} \int r dr \rho_i(r)[\ln\rho_i(r) v_w - 1 + \beta\bar{\mu}_i^\ominus] + \int r dr \langle\rho_p(r)\rangle[f(r)[\ln f(r) + \beta\bar{\mu}_{A^-}^\ominus] + [1 - f(r)]$$

$$\times [\ln(1 - f(r)) + \beta\mu_{AH}^\ominus]] + \sum_{i=\{+,-,H^+,OH^-\}} \beta \int r dr \left(u_i^B(r)\rho_i(r)\right) + \beta \int r dr (u_p^B(r) f(r) \langle\rho_p(r)\rangle)$$

$$+ \beta \int r dr \left[\langle\rho_q(r)\rangle\psi(r) - \frac{1}{2}\epsilon(\nabla_r \psi(r))^2\right]\Biggr\} + \beta\sigma_q \psi(R) + \frac{\beta}{2A_0}\int d^3r d^3r' \chi(|\vec{r} - \vec{r}'|)\langle\rho_p(r)\rangle\langle\rho_p(r')\rangle$$

where $\beta = 1/k_B T_m$ in which $k_B$ is the Boltzmann constant and $T_m$ is the average temperature of the fluid. The first term denotes the conformational entropy of the end tethered polymer brushes in which $P(\alpha)$ is the probability of finding a anchored polymer chain in conformation $\alpha$, and the parameter $\sigma_p = N_p/A_0$ is the grafting density of PE brushes and equals to the ratio of the total number of tethered polymer chains ($N_p$) to the inner surface of the nanopore ($A_0$). The function $P(\alpha)$ may be obtained by the molecular theory approach outlined by Szleifer group [62,63]. We repeat their procedure here for the sake of completeness. The average volume fraction of PE brushes can be evaluated from the probability distribution function as:

$$\langle \phi_p(r) \rangle = \frac{N_p}{A(r)} \sum_\alpha P(\alpha) n_p(\alpha; r) v_p \tag{4}$$

where $A(r)$ is the area of the virtual surface at distance $r$, $v_p$ is the volume of one polymer segment, and $n_p(\alpha; r)dr$ is the number of polymer segments in conformation $\alpha$ within the distance $r$ and $r + dr$. The second term in the expression of $\mathcal{F}$ in Eq. (3) denotes the translational entropy of the solvent and $\rho_w(r)$ is the number density of the water as solvent and $v_w$ is its volume. The volume fraction of the water at $r$ can be expressed as $\phi_w(r) = \rho_w(r) v_w$. The next two terms of the free energy describe the translational entropy and chemical potential of the mobile ionic species. The fifth term in the free energy denotes the free energy associated with the chemical acid-like dissociation equilibrium in which $f(r)$ corresponds to the $r$-dependent fraction of charged monomers and $\langle \rho_p(r) \rangle = \langle \phi_p(r) \rangle / v_p$ is the polymer density at $r$. The first and third terms within the integral denote the entropy of the deprotonated charged state ($A^-$) and protonated state ($AH$), respectively. The second and fourth terms within the integral correspond to the standard chemical potential of the charged and uncharged state, respectively. It is noted that the quantities $\bar{\mu}_{A^-}^\ominus$ and $\bar{\mu}_{H^+}^\ominus$ denote the standard chemical potential for the charged polymer monomer and the proton and do not include the Born energy. Similar to Ref. [62], a bar is used to differentiate these contributions in the chemical potential from each other. The total standard chemical potential of the charged monomer and proton in bulk medium has a Born energy contribution as $\mu_i^\ominus = \bar{\mu}_i^\ominus + u_i^B$ where $u_i^B(r) = e^2 z_i^2/[8\pi \epsilon(r) \mathrm{r}_i]$ is the Born contribution to the free energy of the ionic species in which $\epsilon(r)$ is the $r$-dependent permittivity. Accordingly, $\Delta G^B = u_i^B(r) - u_{i,free\ space}^B$ is, following Born, the change in free energy of charge transfer from the free space to a medium with permittivity $\epsilon$. Hence, the total electrostatic self-energy of all charges present in an inhomogeneous medium through the nanopore can be expressed as:

$$u^B = \sum_i \int 2\pi r dr u_i^B(r) \rho_i(r) \tag{5}$$

In order to describe the inhomogeneous medium through the soft nanopore and to calculate the Born energy contribution to the free energy via seventh and eighth terms of $\mathcal{F}$, the following constitutive equation is employed to model the permittivity function $\epsilon(r)$ as the average of the permittivity functions of the components of the system weighted by their respective local volume fractions [62] as $\epsilon_{eff}(r) = \epsilon_p\langle\phi_p(r)\rangle + \epsilon_w(1 - \langle\phi_p(r)\rangle)$. Besides, an alternative approximation has been proposed by Croze and Cates [64] which is based on the application of the Maxwell–Garnett mixing formula as:

$$\epsilon_{eff}(r) = \epsilon_w \left( \frac{3\gamma^3(2 + \Lambda) - 8\pi(1 - \Lambda)\langle\phi_p(r)\rangle}{3\gamma^3(2 + \Lambda) + 4\pi(1 - \Lambda)\langle\phi_p(r)\rangle} \right) \tag{6}$$

where $\Lambda = \epsilon_p/\epsilon_w$ and $\gamma$ is a correction factor introduced by Croze and Cates [64] to account for the difference between the electrostatic and volume size of the polymer. The numerical value of this factor may be chosen in accordance with Croze and Cates [64] as $\gamma = 1.5$. The above expression gives the effective electrostatic permittivity of a random mixture of spherical inclusions of permittivity $\epsilon_p$ residing in a medium of permittivity $\epsilon_w$ [63]. The permittivity of the polymer and that of the solvent and ions is taken as $\epsilon_p = 2\epsilon_0$ and $\epsilon_w = 78.5\epsilon_0$, respectively, in which $\epsilon_0$ is the permittivity of the vacuum.

The ninth and tenth terms in Eq. (3) denote the electrostatic contribution to the $\mathcal{F}$. Here, $\psi(r)$ corresponds to the electrostatic potential at position $r$, and $\sigma_q$ is the surface charge density. Also, $\langle\rho_q(r)\rangle$ is the local charge density at $r$ which equals to:

$$\langle\rho_q(r)\rangle = -ef(r)\langle\rho_p(r)\rangle + ez(\rho_+(r) - \rho_-(r)) + e(\rho_{H^+}(r) - \rho_{OH^-}(r)) \tag{7}$$

Non-electrostatic van der Waals interactions are represented by the last term of the free energy functional. It describes the solvent quality. Here $\chi(|\vec{r} - \vec{r}'|) = -\varepsilon_{vdW}\ell^6/|\vec{r} - \vec{r}'|^6$ is an attractive van der Waals potential, where $\ell$ denotes the segment length of the polymer [62]. The variable $\varepsilon_{vdW}$ describes the strength of the van der Waals interactions. As noted by Ref. [62], for a good solvent $\chi(|\vec{r} - \vec{r}'|) = 0$.

The intermolecular excluded volume interactions are accounted for by the following constraint assuming that the system is incompressible at every position [62]:

$$\langle\phi_p(r)\rangle + \sum_{i=\{w,+,-,H^+,OH^-\}} \phi_i(r) = 1 \tag{8}$$

As noted by Ref. [62], these volume constraints are enforced through the introduction of the Lagrange multipliers $\pi(z)$. It is reminded from Ref. [62] that the volume fraction of each species explicitly encompasses its size, and for polymers, includes a sum over the possible conformations. The free energy is minimized with respect to $P(\alpha)$, $\rho_i(r)$, and $f(r)$, and varied with respect to the electrostatic potential $\psi(r)$ under the constraints of incompressibility and the fact that the system is in contact with a bath of cations, anions, protons, and hydroxyl ions [62]. The proper thermodynamic potential can be considered as[62]:

$$w = \frac{\beta \mathcal{F}}{A_0} - \frac{\beta}{R} \sum_{i=\{w,+,-,H^+,OH^-\}} \mu_i \int r dr\, \rho_i(r) - \frac{\beta}{R}\mu_{H^+} \int r dr\, (1 - f(r))\langle\rho_p(r)\rangle$$
$$+ \frac{\beta}{R} \int r dr\, \pi(r)[\langle\Phi_p(r)\rangle + \sum_{i=\{w,+,-,H^+,OH^-\}} \phi_i(r) - 1] \tag{9}$$

Also, the probability distribution function is given by the following expression [62]:

$$P(\alpha) = \frac{1}{q}\left\{\exp\left[-\beta\int dr n_p(\alpha;r)\left\{\pi(r)v_p + \psi(r)(ez) + \frac{1}{\beta}\ln f(r) - \frac{1}{2}\frac{\delta\epsilon(r)}{\delta\rho_p(r)}\left(\frac{\partial\psi(r)}{\partial r}\right)^2\right\}\right]\right.$$
$$\times \exp\left[-\beta\int dr dr' n_p(\alpha;r)A(r)\chi(|\vec{r}-\vec{r}'|)\langle\rho_p(r')\rangle\right] \quad (10)$$
$$\left.\times \exp\left[-\beta\int dr n_p(\alpha;r)\left(u_p^B(r) - \sum_i{}' u_i^B(r)\rho_i(r)\frac{\delta\ln\epsilon(r)}{\delta\rho_p(r)}\right)\right]\right\}$$

where $q$ is a normalization factor ensuring that $\sum_\alpha P(\alpha) = 1$. The first term in the exponential results from the repulsive excluded volume interactions that a polymer in conformation $\alpha$ possesses [62]. The second and fourth terms include a purely electrostatic contribution whereas the third term is an entropy like term associated with the degree of charging of the chargeable group [62]. The fifth term denotes the van der Waals interaction among the monomers. The last two terms originate from the coupling of the polymer density to the Born self-energy [62]. In the final term of the probability distribution function, the summation over $i$ runs over all charged species, *i.e.*, the co- and counter ions, protons, hydroxyl ions, and the charged groups in the PEL, and, thus, $i = p$ involves $f(r)\langle\rho_p(r)\rangle$. This fact is indicated here and in the following equations through the prime in the summation. To see further discussion about this equation, the readers are referred to study of Nap *et al.* [62].

The local volume fraction of the mobile ions and that of the solvent can be, respectively, written as:

$$\rho_i(r)v_w = \exp\left(\beta(\mu_i - \bar{\mu}_i^\ominus) - \beta\pi(r)v_i - \beta ez_i\psi(r) + \frac{1}{2}\beta\frac{\delta\epsilon(r)}{\delta\rho_i(r)}\left(\frac{\partial\psi(r)}{\partial r}\right)^2\right)$$
$$\times \exp\left(-\beta u_i^B(r) + \beta\sum_j{}' u_j^B(r)\rho_j(r)\frac{\delta\ln\epsilon(r)}{\delta\rho_j(r)}\right) \quad (11)$$

$$\rho_w(r)v_w = \exp\left(-\beta\pi(r)v_w + \frac{1}{2}\beta\frac{\delta\epsilon(r)}{\delta\rho_w(r)}\left(\frac{\partial\psi(r)}{\partial r}\right)^2 + \beta\sum_i{}' u_i^B(r)\rho_i(r)\frac{\delta\ln\epsilon(r)}{\delta\rho_w(r)}\right) \quad (12)$$

Considering the case where the permittivity function only explicitly depends on the polymer density, the expression for the local volume fraction of the mobile ions and that of the solvent can be simplified as:

$$\rho_w(r)v_w = \exp(-\beta\pi(r)v_w) \quad (13)$$

$$\rho_i(r)v_w = \exp\left(\beta(\mu_i - \bar{\mu}_i^\ominus) - \beta\pi(r)v_i - \beta ez_i\psi(r) - \beta u_i^B(r)\right) \quad (14)$$

Similar to the discussion of Ref. [62], based on the incompressibility constraint, charge neutrality and the water self-dissociation equilibrium in the bulk salt solution, the local volume fraction of the mobile ions can be readily expressed by relating them to their bulk concentrations ($C_i^{bulk}$) as:

$$\rho_i(r)v_w = \rho_i^{bulk}v_w\exp\left(-\beta[\pi(r) - \pi^{bulk}]v_i - \beta ez_i\psi(r) - \beta\left(u_i^B(r) - u_{i,bulk}^B\right)\right) \quad (15)$$

In this regard, the bulk concentrations the protons and hydroxyl ions can be expressed, respectively, as $C_{H^+}^{bulk} = 10^{-\text{pH}}$ and $C_{OH^-}^{bulk} = 10^{\text{pH}-14}$, and for the remaining cations and anions in the bulk salt solution, we may write:

$$C_+^{bulk} = C_R, \quad C_-^{bulk} = C_R + C_{H^+}^{bulk} - C_{OH^-}^{bulk} \; ; \quad \text{pH} \leq 7 \tag{16}$$

$$C_+^{bulk} = C_R - C_{H^+}^{bulk} + C_{OH^-}^{bulk}, \quad C_-^{bulk} = C_R \; ; \quad \text{pH} > 7 \tag{17}$$

where $C_R$ is the bulk concentration of salt solution. Then, minimizing the free energy functional with respect to the fraction of charged monomers leads to:

$$f(r) = \frac{K_a^{\ominus} e^{-\beta \Delta G_s(r)} \exp(-\beta \pi(r) v_{H^+})}{\rho_{H^+}(r) v_{H^+} + K_a^{\ominus} e^{-\beta \Delta G_s(r)} \exp(-\beta \pi(r) v_{H^+})} \tag{18}$$

where $K_a^{\ominus} = \exp(-\beta \Delta G_a^{\ominus})$ is the chemical equilibrium constant in which $\Delta G_a^{\ominus} = \mu_{A^-}^{\ominus} + \mu_{H^+}^{\ominus} - \mu_{HA}^{\ominus}$ is the standard free energy change of the reaction $HA \leftrightharpoons A^- + H^+$. In this regard, $K_a^{\ominus}$ can be related to the corresponding experimental equilibrium constant of a single acid molecule in dilute solution as $K_a(r) = CK_a^{\ominus}$ in which the constant $C$ is introduced for consistency of units. Also, $\Delta G_s(r) = \Delta u_p^B(r) + \Delta u_{H^+}^B(r)$ is the difference in self-energy of creating a charged polymer monomer and proton pair at position $r$ and that in the bulk solution. In the expression of $\Delta G_s(r)$, the value of $\Delta u_i^B(r)$ is the difference between the Born energy of species $i$ at position $r$ and its value in bulk solution [62].

The variation of $\mathcal{F}$ with respect to the electric potential yields the Poisson equation as follows:

$$\frac{1}{\bar{r}} \frac{d}{d\bar{r}} \left( \bar{r} \bar{\epsilon}(r) \frac{d\bar{\psi}(r)}{d\bar{r}} \right) = -K^2 \langle \bar{\rho}_q(r) \rangle \tag{19}$$

where $\bar{r} = r/\ell$ is the dimensionless radial coordinate scaled by the segment length of the PE brushes ($\ell$), $\bar{\epsilon} = \epsilon_{eff}/\epsilon_w$, $\bar{\psi} = e\beta\psi$, $K = \ell/\lambda_D$ and $\bar{\rho}_q = \rho_q/(ez\rho_R)$. In these parameters, $\rho_R$ is the ionic number density in the bulk salt solution, and $\lambda_D = (\rho_R e^2 z/\epsilon_w k_B T_m)^{-1/2}$ is the Debye length which characterizes the thickness of EDL or equivalently the bulk concentration of electrolyte solution for constant radius of the nanopore. The boundary conditions of Eq. (19) are defined as: (a) the axial symmetry condition for $\bar{\psi}$, and (b) the gauss law for surface charge density at the wall of the nanopore which can be expressed, respectively, as:

$$(a) \; \left. \frac{d\bar{\psi}(r)}{d\bar{r}} \right|_{\bar{r}=0} = 0, \quad (b) \; \bar{\epsilon} \left. \frac{d\bar{\psi}(r)}{d\bar{r}} \right|_{\bar{r}_s} = \bar{\sigma}_q \tag{20a, b}$$

where $\bar{r}_s = R/\ell$ and $\bar{\sigma}_q = \sigma_q \ell e\beta/\epsilon_w$. The procedure introduced by [62] is employed here to evaluate the charge distribution and other properties of PEL inside the soft nanotube. As discussed by Ref. [62], the unknowns in Eqs. (10), (13), (15), (18) and (19) are the Lagrange multiplier or radial pressure, $\pi(r)$, the electrostatic potential, $\psi(r)$, and the polymer volume fraction, $\langle \rho_p(r) \rangle$. Application of the molecular theory requires the evaluation of these variables. This can be accomplished by substituting the expressions of the probability distribution function of chain conformations and the volume fractions of all components into the incompressibility constraints and the Poisson equation. Combining these two equations with the equation for the polymer volume fraction, Eq. (4), results in a set of integro-differential equations whose solution will determine the radial pressure, electrostatic potential, the degree of protonation, and the volume fraction of the PEL, the ions, and the solvent [62]. It is noted that a similar fashion exists for the minimization of the free energy functional of a system including the PB grafted nanopore, which is not presented here for the sake of brevity.

## 4. Calculation of velocity field considering the slip boundary condition

Having obtained the electric charge distribution inside the nanopore, we may proceed to evaluate the electric charge density, by which, the electroosmotic body force, and also velocity field, in the Navier–Stokes–Brinkman equation (2) can be calculated. For the present problem, the axial electric field $E_z$ in Eq. (2) may be generated by the external electrodes placed at the two ends of the nanopore. The dimensionless form of NSB equation (2) for axisymmetric EOF velocity can be expressed in the cylindrical coordinates as:

$$\frac{d^2\bar{u}(\bar{r})}{d\bar{r}^2} + \frac{1}{\bar{r}}\frac{d\bar{u}(\bar{r})}{d\bar{r}} - K^2\big(\mathbb{z}[\bar{\rho}_+(\bar{r}) - \bar{\rho}_-(\bar{r})] + \bar{\rho}_{H^+}(\bar{r}) - \bar{\rho}_{OH^-}(\bar{r})\big)\bar{E}_z - \alpha^2\langle\phi_p(\bar{r})\rangle^2\bar{u} = 0 \quad (21)$$

In above equation, the dimensionless quantities are defined as follows: $\bar{E}_z = E_z/E_0$ is the dimensionless applied electric field, $\bar{\rho}_i = \rho_i/\rho_R$ is the dimensionless number density of ionic species, $\alpha = \ell/a$ is the ratio of segment length of the PE brushes to the effective monomer size [65], and $\bar{u} = u/u_{HS}$ is the dimensionless velocity. In these quantities, $u_{HS} = -\epsilon_w k_B T_m E_0/(e\mathbb{z}\eta)$ is the Helmholtz-Smoluchowski velocity [66, 67] and $E_0$ is the characteristic electric field. The boundary conditions for Eq. (21) are defined as: (a) the axial symmetry condition for $\bar{u}$, and (b) the slip velocity condition at the wall of the nanotube which can be expressed, respectively, in the following dimensionless forms:

$$(a)\ \frac{d\bar{u}}{d\bar{r}}\bigg|_{\bar{r}=0} = 0, \quad (b)\ \bar{u}|_{\bar{r}_s} = -\bar{\mathcal{L}}_s\frac{d\bar{u}}{d\bar{r}}\bigg|_{\bar{r}_s} \quad (22a, b)$$

where $\bar{\mathcal{L}}_s = \mathcal{L}_s/\ell$ is the dimensionless Navier slip coefficient where, depending on the roughness and structure of the surfaces and the size of the nanopore, may vary from sub-nanometers up to several micrometers [68-70]. For $\bar{\mathcal{L}}_s = 0$, the problem is simplified to the case of no-slip boundary condition.

## 5. Numerical simulation of electric potential and velocity fields

At first, in order to obtain the properties of PEL and charge distribution through the soft nanopore, we need to generate a representative set of polymer conformations. In this regard, we use the Rotational Isomeric State (RIS) model [55] for the generation of polymer chains with a segment length of $\ell = 0.5$ nm. All generated conformations are self-avoiding and do not penetrate the tethering surface. Further details on the chain modeling can be found in Refs. [62, 63, 71, 72].

Then, we convert the aforementioned differential equations into a set of coupled nonlinear equations by space discretization. For this purpose, we discretize the differential equations of the present problem by utilizing a central finite difference method with second order spatial precision, and the resulting set of algebraic equations are solved numerically. The inputs required to solve the nonlinear equations are the bulk pH, salt concentration, a set of polymer conformations, the surface coverage of the end-tethered PE brushes, the $pK_a$ or $pK_b$ of the acidic or basic groups, the Born radii of all charged species, the volume of all species, and the Navier slip coefficient [62]. In accordance with Ref. [62], the corresponding properties of various molecular species are given below in Table 1.

Table 1: The physical and electrical properties of various molecular species

| Molecular species | $v$ (nm³) | $z$ | $r$ |
|---|---|---|---|
| Polymer | 0.113 | −1 | 0.3 |
| Water | 0.03 | 0 | 0.193 |
| $H^+, OH^-$ | 0.03 | +1, −1 | 0.193 |
| +, − | 0.034 | +1, −1 | 0.2 |

## 6. Parameters quantifying the characteristics of PEL and EOF

In this section, we define the main parameters quantifying the characteristics of PEL and EOF, namely, (1) *ionic conductance*, (2) *ionic selectivity*, (3) *the average degree of PEL charge*, and (4) *the volumetric flow rate* inside the nanopore. By defining the net ionic current ($I_t$) as the sum of the advection and conduction currents, we may write [36]:

$$I_t = (G_{adv} + G_{cond})V = \sum_{i=\{^{+,-,}_{OH^-,H^+}\}} I_i = 2\pi FV \sum_{i=\{^{+,-,}_{OH^-,H^+}\}} \left\{ \frac{z_i}{V} \int_0^R C_i(r)u(r)rdr + \frac{Fz_i^2 D_i}{RTL} \int_0^R rdrC_i(r) \right\} \quad (23)$$

where $L$ is the length of the nanopore, $F$ is the Faraday constant, $z_i$ and $D_i$ are the charge and diffusion coefficient of the ion $i$, respectively, and $C_i(r)$ is the concentration of the ion $i$ at $r$ position in equilibrium conditions, which is determined from the molecular theory approach. Also, $V$ is the applied electric field, and $G_{adv}$ and $G_{cond}$ are the advective and conductive parts of the ionic current per unit applied electric potential, respectively. Specifically, $G_{cond}$ can be termed as the conductivity of the nanopore [18, 62]. Also, to quantify the overall charge of the PEL, the average degree of charge of the PEL can be evaluated as [62]:

$$\langle f \rangle = \frac{\int rdr\, f(r)\langle \rho_p(r) \rangle}{\int rdr\, \langle \rho_p(r) \rangle} \quad (24)$$

Moreover, we may define the ionic selectivity of the nanopore as the ratio of the difference in the net currents of ions of opposite signs (for example, $I_- - I_+$ or vise versa) to the total current carried by ions of salt solution [73]. Therefore, we may express the ionic selectivity of the nanopore, $I_s$, as follows:

$$I_S = \frac{\left|\sum_{i=\{-,OH^-\}} I_i\right| - \left|\sum_{i=\{+,H^+\}} I_i\right|}{\left|\sum_{i=\{-,OH^-\}} I_i\right| + \left|\sum_{i=\{+,H^+\}} I_i\right|} \quad (25)$$

where $I_i$ is defined through Eq. (23). Based on this definition, the positive and negative signs of the ionic selectivity, $I_S$, imply an anion and cation selective nanopore, and the zero ionic selectivity corresponds to a non-selective case.

Another important parameter is the volumetric flow rate ($Q$) which can be expressed in the dimensionless form as follows:

$$\bar{Q} = Q/(\pi R^2 u_{HS}) = 2\ell^2/(R^2) \left| \int \bar{r} \mathrm{d}\bar{r} \bar{u} \right| \tag{26}$$

where the integral in above equation is taken over the entire inner surface of the nanopore, and can be evaluated by the Simpson's 3/8 rule.

## 7. Results and discussion

The main objective of the present study is the quantitative investigation of the electrokinetic characteristics of solid-state soft nanopore by presenting the results of molecular theory study of EOF in the PE-grafted nanopore. Also, accounting for the influence of interfacial slip and conformation-dependent permittivity of the medium, we aim to provide a more comprehensive information regarding the role of soft interface on the distribution of ions, potential and velocity fields, *etc*. Therefore, the present problem incorporates all factors controlling the pH of the electrolyte, bulk concentration of the salt solution ($C_{KCl}$), grafting density of the end tethered polymer brushes ($\sigma_p$), equilibrium constant of the ionization reaction of the acidic or basic groups (denoted as $pK_a$ or $pK_b$, respectively), amount of fluid/wall slippage ($\bar{\mathcal{L}}_s$), Temperature of the fluid ($T_m$), radius ($R$) and length ($L$) of the nanopore, permittivity of the polymer brushes ($\epsilon_p$) and other intrinsic properties of monomers of the PEL and ionic species of the electrolyte. Such number of controlling parameters provide a high degree of flexibility in terms of fine tuning the ions transport process, and consequently, allow the design of more efficient nanofluidic devices.

At first, in order to check the validity of our two approaches and to compare the present results with previous experimental studies, we include the data of Ref. [18] in Fig. 2(a) to show the pH dependency of conductivity of soft nanopore used in the experiment. The radius and length of the utilized cylindrical nanopore were $R = 7.5$ nm and $L = 12$ $\mu$m, respectively, and was modified with an end tethered layer of poly(4-vinyl pyridine), 4PVP. The nanopore was connected to two reservoirs containing aqueous 0.1M KCl solution. The mobile ions in the system ($K^+$, $H^+$, $OH^-$ and $Cl^-$) flow through the pore upon applying an electric potential difference between the electrodes located at the reservoirs. Each segment in 4PVP can be either positively charged (protonated) or neutral, depending on the values of $pK_a$ and the pH [62]. Due to the presence of a methylene group in the para position to the pyridinic N, the value of $pK_a$ for 4PVP is expected to be in the range from 5.23 to 5.98 which the extrema correspond to the $pK_a$ of pyridine and that of 4-ethyl pyridine [62]. Therefore, we use $pK_a = 5.4$ for the determining the conductivity of soft nanopore.

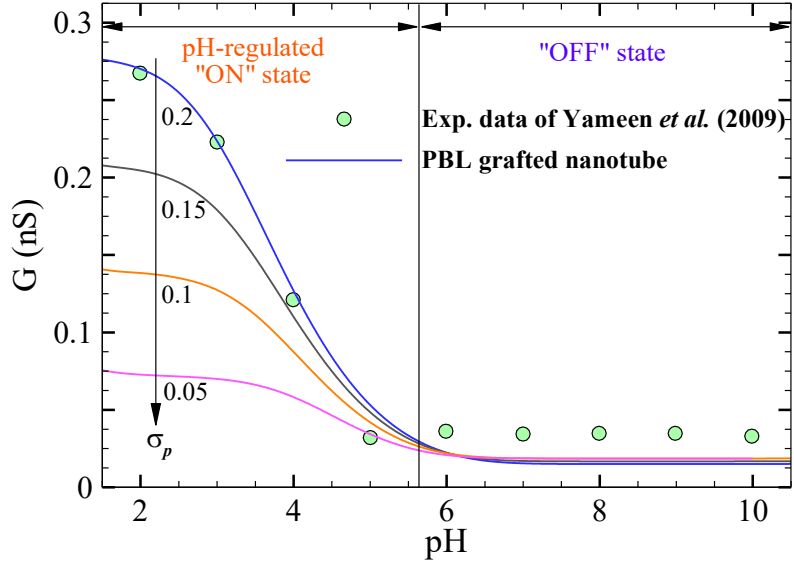

**Fig. 2:** Variation of ionic conductivity of the nanopore with pH of the salt solution in various densities of tethered polymer chains per unit area. The symbols (○) correspond to the experimental study of Yameen *et al*. [18].

As can be seen from Fig. 2(a), by using the chain length of 28, $\sigma_p = 0.2$ nm$^{-2}$ and $pK_a = 5.4$ for the PB bushes, the results of the present study are in accord with those of Ref. [18] demonstrating the reliability of the molecular theory approach. Also, by extending the results of Yameen *et al*. [18] shown in Fig. 2(a) for various values of $\sigma_p = 0.05, 0.1$ and $0.15$ at $pK_a = 5.4$, we conclude that the ionic conductivity of solid-state nanopore decreases with $\sigma_p$ for pH $\leq 6$, while increases for higher values of pH. Such variation indicates this fact that, due to the basic nature of the PE-brushes, increasing the polymer grafting density simultaneously increases the overall degree of polymerization leading to a higher concentration of counterions inside the nanopore. While, increasing $\sigma_p$ in the range pH > 6 results in the rejection of ions from the nanopore due to volume exclusion effect of macromolecules. In Fig. 2, the "ON" state shows the case in which the mobile ionic species can be conducted by the nanopore, while in the other case, the nanopore is switched off indicating a very low values of its conductivity.

In the present analysis, two different cases of poly-acid layer (PAL) and poly-base layer (PBL) are considered for the nature of soft layer grafted to the inner surface of the nanopore. Through the analysis, the set of default parameters are taken as pH = 5, $pK_a = 5$, $pK_b = 9$, $\alpha = 1$, $\mathcal{L}_s = 0.25$ nm, $\sigma_p = 0.1$ nm$^{-2}$, $C_{KCl} = 100$ mM and $\bar{\sigma}_q = 0$, and from now on, if we do not restate the value of a parameter, it is assumed that the above reference value is adopted for that parameter. Regarding the ionization constant of the PEL, the typical value of $pK_a = 5$ is comparable to that of a carboxylic acid group, i.e., polyacrylic acid, and the typical value of $pK_b = 9$ is approximately close to that of 4PVP (~8.7) [62]. Also, the radius and the length of the nanotube is taken as $R = 20$ nm and $L = 12$ $\mu$m [18], and each polymer chain is considered to have 40 monomers with the segment length of 0.5 nm.

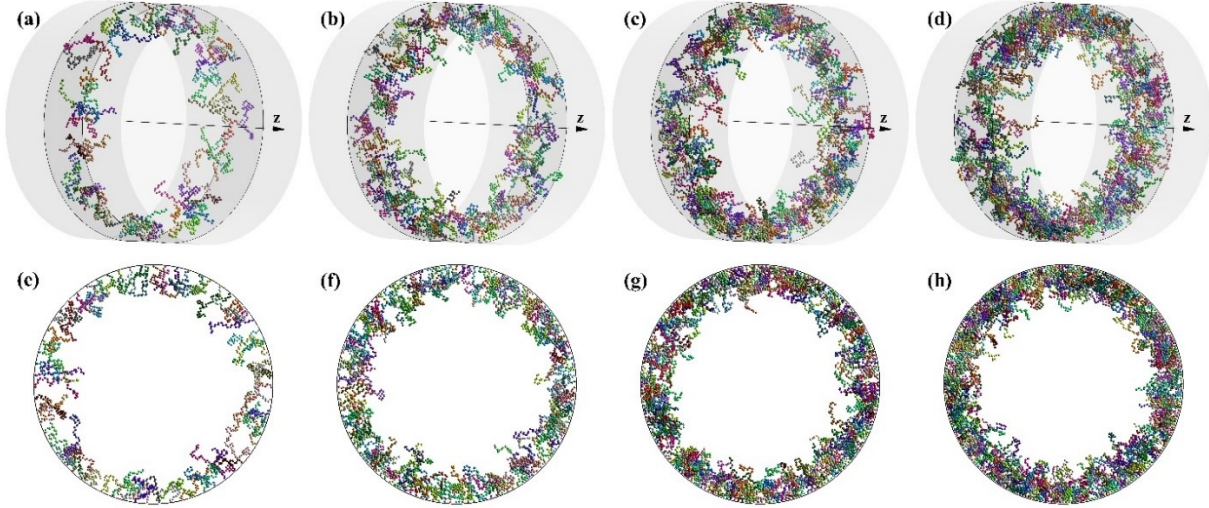

**Fig. 3:** 2D and 3D illustration of representative distribution of PE brushes (distinguished from each other by different colors) having the length of 40 segments inside the nanopore at various $\sigma_p =$ **(a, e)** 0.05 nm$^{-2}$ **(b, f)** 0.1 nm$^{-2}$, **(c, g)** 0.15 nm$^{-2}$, and **(d, h)** 0.2 nm$^{-2}$. The selected chains are at a cross slice of width 10 nm around $z = 0$.

Under the aforementioned conditions, a representative distribution of PE brushes having the chain length of 40 segments inside the nanopore at various $\sigma_p = 0.05, 0.1, 0.15$ and $0.2$ nm$^{-2}$ is shown in Fig. 3, in which a bunch of polymer chains are randomly tethered to the surface of a cross slice of width 10 nm around $z = 0$, i.e., $z \in [-5 \text{ nm}, 5 \text{ nm}]$. All PE-chains illustrated in Fig. 3 are generated by using the rotational isomeric states of $t$, $g^+$ and $g^-$ ($trans$, $gauche^+$, $gauche^-$) [55], in which, all bonds are of the same length. Through the modeling of polymer chain conformations, the totally placement of PE-brush inside the nanopore has been checked. As can be seen from Fig. 3, for $\sigma_p = 0.2$ nm$^{-2}$, the macromolecules occupy a large amount of space near the wall of the nanopore at room temperature. The represented features can help us to gain a physical insight into the role of polymer chains in decreasing the volume fractions of mobile species near and far from the wall. We may use this figure for explaining the role of grafting density of the PE brushes in the alteration of conformation-dependent properties of EOF in soft nanopore.

Then, we show that how the physio-chemical properties of polymer chains affect the electrostatic and hydrodynamic characteristics of electroosmotic flow through soft nanotube. Figures 4(a-f) represent the volume fraction of polymer and water molecules and the effective relative permittivity ($\epsilon_{r,eff} = \epsilon_{eff}/\epsilon_0$) for different grafting density of polymer chains, $\sigma_p = 0.05, 0.1$ and $0.2$ for two cases of nanopore grafted with PAL and PBL. As can be seen, by increasing the number of tethered polyelectrolytes per unit area, the polymer density increases resulting in the dropping the amount of volume fraction of water molecules and effective permittivity in the wall adherent regions. Also, it can be seen from Figs. 4(a, d) that the distribution of the polymer segments in soft nanopore is approximately identical for both PAL and PBL under the same parametric conditions. In other words, the nature of the PEL does not make any significant structural change in the physical distribution of monomers within this layer. However, we will show that electrochemical properties of the whole medium strongly depend on the type of the PE brushes.

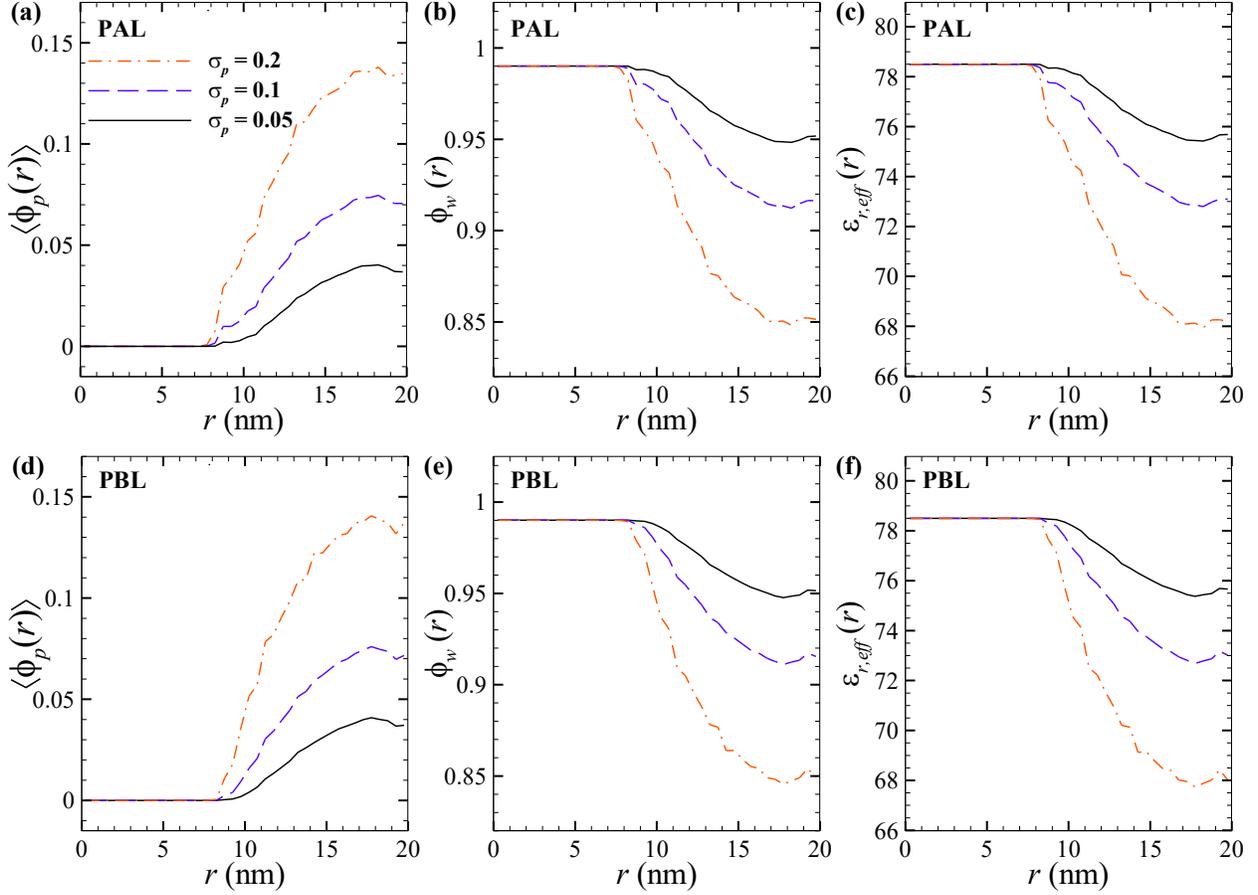

**Fig. 4:** Radial distribution of **(a, d)** polymer volume fraction, **(b, e)** volume fraction of water molecules, and **(c, f)** effective relative permittivity of medium at different values of polymer grafting density. All panels are in the same parametric conditions, and top/bottom panels correspond to the case of nanopore grafted with PAL/PBL.

Figures 5(a-f) represents the radial distribution of ionic species in soft nanopore for two cases of PA and PB grafted nanopores. As can be seen, the type of the polymer chains determines the concentration of each ionic species in soft nanopore, where for the case of nanopore grafted with PAL, the concentration of cations is significantly larger than that of anions. It is noted that the concentration of $OH^-$ can be evaluated by the relation of pOH $= 14 -$ pH, such that by using this equation, the radial distribution of $OH^-$ ions can be determined using the data of Fig. 5(a). The accumulation of positive ions in the region of the PEL is due to the fact that the PAL undergoes the dissociation ionization reaction resulting in the formation of positive sites within the PEL by which the negative ions are impelled through this layer. The opposite condition holds for PBL, so that as can be seen in Figs. 5(d-f), the concentration of negatively charged ions decrease in near wall regions. Moreover, it noteworthy that, increasing the grafting density of PEL results in the intensification of the repulsive interaction of coions with charged sites within the polymeric layer, such that, for $\sigma_p = 0.2$ nm$^{-2}$, the overall concentration of $K^+$ in PA grafted nanopore is as large as ~200 mM, while the mean concentration of $Cl^-$ is as small as ~40 mM providing appropriate conditions for the separation of ions with opposite charge.

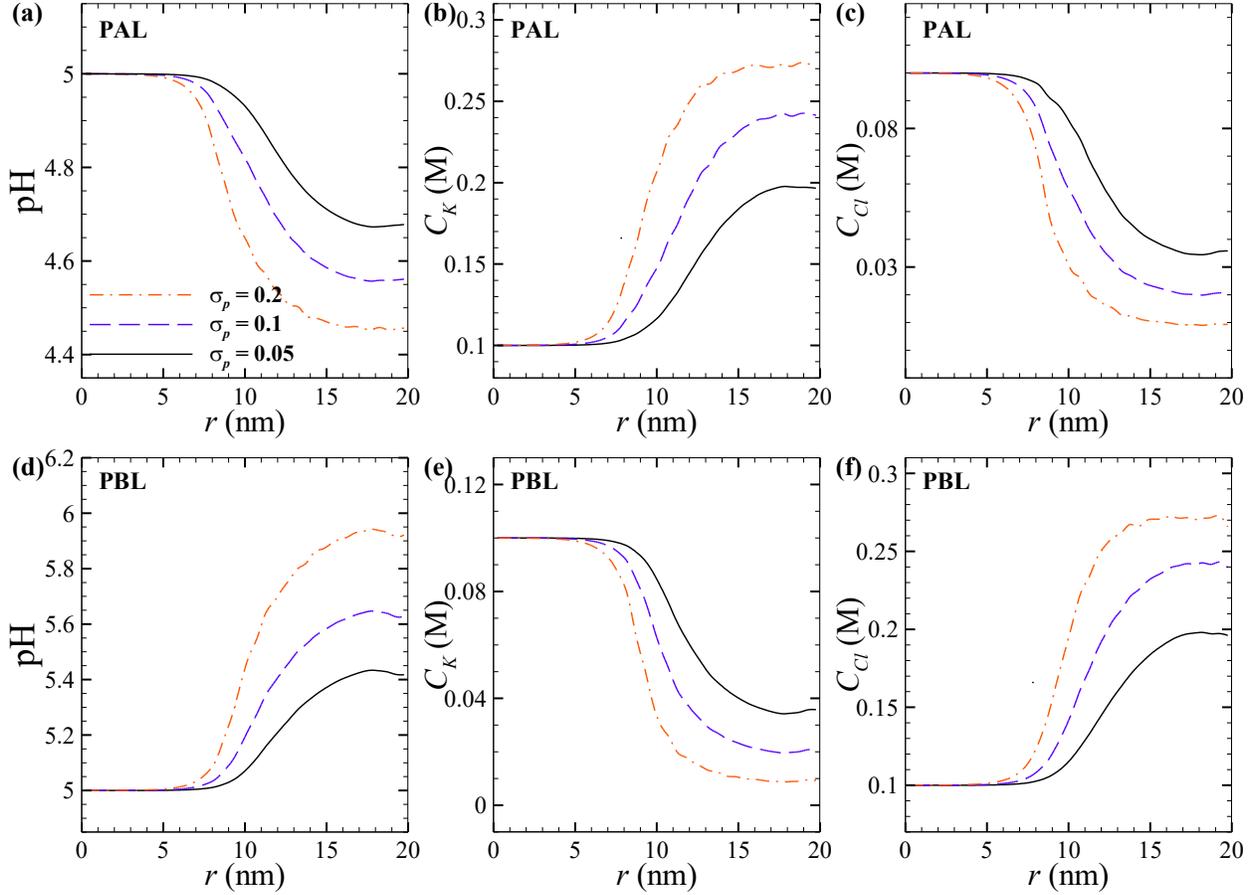

**Fig. 5:** Radial variation of **(a, d)** the local pH of the salt solution, **(b, e)** molar concentration of potassium ions, and **(c, f)** molar concentration of chloride ions at different values of polymer grafting density. All panels are in the same parametric conditions, and top/bottom panels correspond to the case of nanopore grafted with PAL/ PBL.

In order to further investigate the electrostatic interaction between the electrolyte solution and polymer layer grafted to the nanopore, we examine the change in the radial distribution of electric potential and the local degree of charge of the polymer chains for various salt concentrations, and the corresponding results are shown in Figs. 6(a, b, d, e). As can be seen, for both cases of nanopore grafted with PAL and PBL, increasing the ionic strength of the electrolyte decreases the magnitude of the average electric potential inside the soft nanopore. This behavior can be attributed to the fact that the electrostatic screening of the charges of the PEL by electrolyte solution is less provided by the dilute salt solutions. Moreover, the degree of charge of the polymer brushes is strongly affected by the bulk concentration of electrolyte solution, such that, by increasing the value of $C_{KCl}$, the degree of polymer charge in both cases of PA and PB grafted nanopore is increased. This phenomenon is caused by decreasing the volume fraction of $H^+$ or $OH^-$ ions in the PEL region, by which, the dissociation or association reaction of polymer chains is accelerated in favor of the charged state of the monomers, and therefore, the number of charged sites within the PEL is increased.

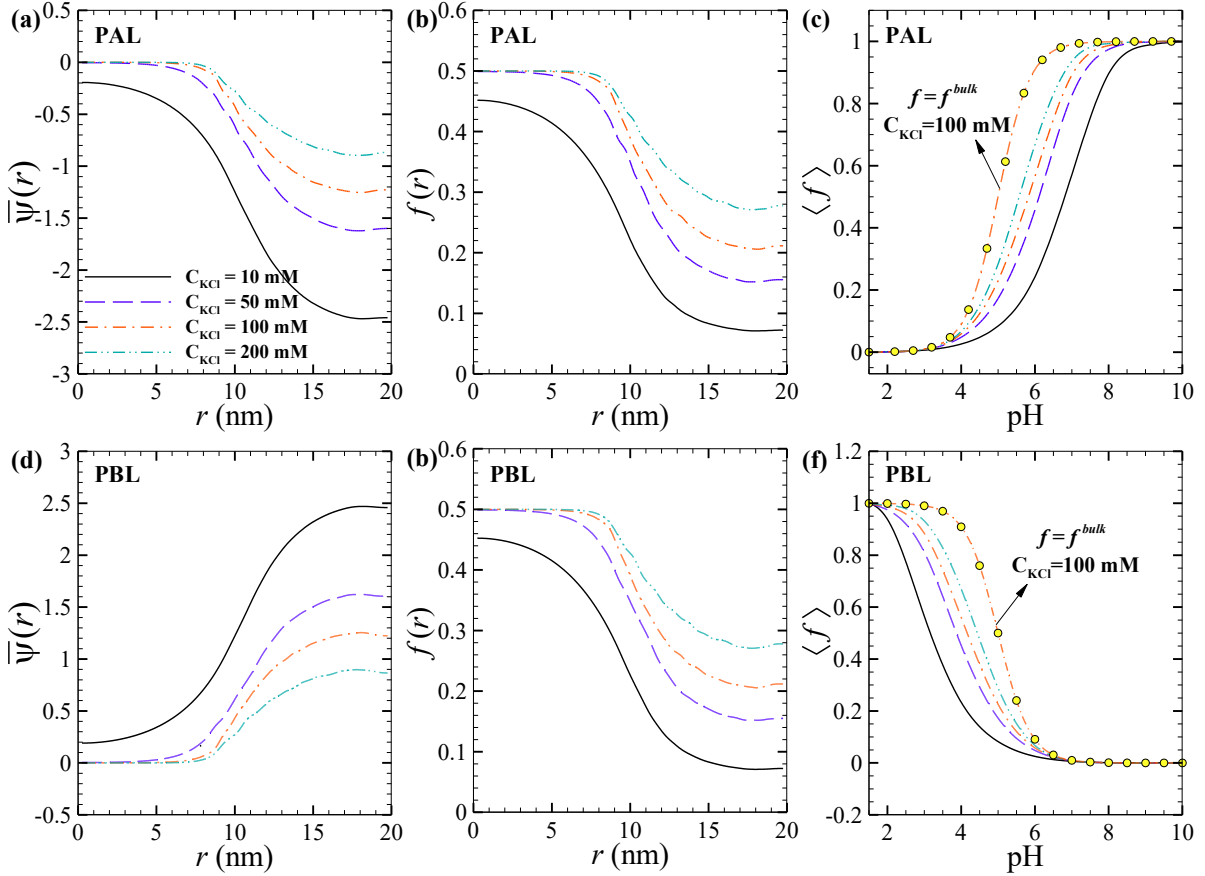

**Fig. 6:** Radial distribution of **(a, d)** dimensionless electric potential, **(b, e)** degree of polymer charge, and **(c, f)** the variation of average degree of polymer charge as a function of pH at different salt concentrations. All panels are in the same parametric conditions, and top/bottom panels correspond to the case of nanopore grafted with PAL/ PBL.

The effect of pH of the salt solution on the average degree of charge of the polymer brushes for two cases of nanopore grafted with PAL and PBL is also included in Figs. 6(c, f). As can be seen, increasing the pH results in an increase/decrease of the degree of the charged sites within the PA/PB brushes, while increasing the salt concentration results in an increase of $\langle f \rangle$ in both types of the PELs. The reason of this phenomenon can be explained by the *Le Chatelier* principle [63] as follows; by increasing the salt concentration, the number of protons (or hydroxyl ions) inside the nanopore decreases, and therefore, the number of dissociated *AH* groups (or associated *B* groups) increases in order to compensate for the reduction of the number of counterions inside the nanopore. Also, based on this principle, the observed increasing/decreasing behaviors of $\langle f \rangle$ for PAL/PBL with respect to pH is due to acceleration of ionization reaction of PA/PB brushes in favor of charged/uncharged states of the monomers. Moreover, for indicating the importance of considering the local variation of pH, the results of the degree of charge of a single acid/base molecule in the bulk following the ideal chemical equilibrium equation: $f^{bulk}/[1-f^{bulk}] = pK_a/pH$ and $f^{bulk}/[1-f^{bulk}] = pK_b/pOH$ for acid and base molecules are shown in Figs. 6(c, f) by line with symbols. It can be seen from this figure that, ignoring the local pH-regulation of PE-molecules leads to a degree of inaccuracy in the conductivity results specially at intermediate pH intervals.

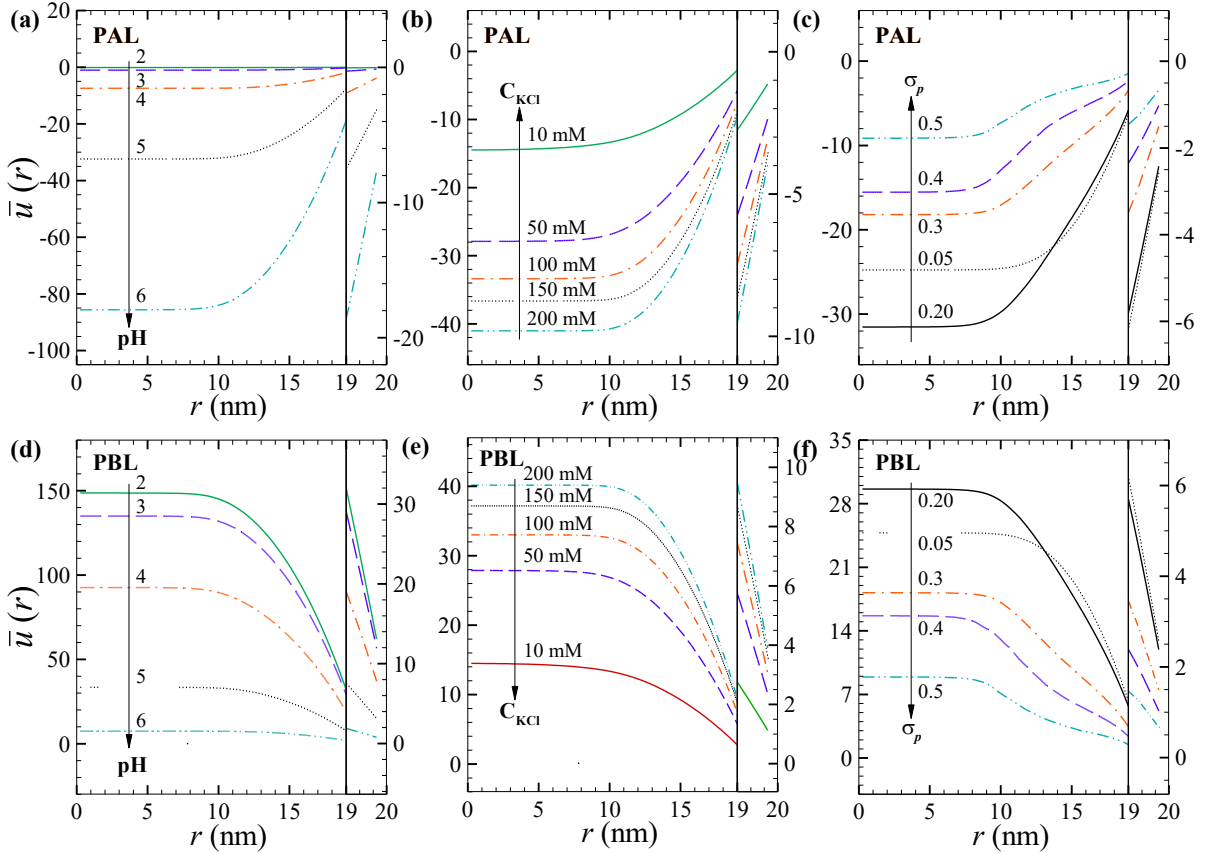

**Fig. 7:** Velocity profiles for various values of **(a, d)** pH of the electrolyte, **(b, e)** salt concentrations, and **(c, f)** the grafting density of PE-brushes. The top/bottom panels correspond to the case of nanopore grafted with PAL/ PBL.

Figures 7(a, d) illustrate the radial distribution of velocity profile for different values of pH of the salt solution. As can be seen, increasing the pH of the electrolyte results in the rising of the velocity plateaus through the nanopore grafted with PAL, while the reverse dependence of velocity profile on pH exists for the case of PB grafted nanopore. Such dual behavior is attributed to the acceleration/deceleration of ionization equilibrium reaction of PAL/PBL via increasing the pH of the electrolyte which in turn results in an increase/decrease of electroosmotic body force on the fluid. It is noted that, for more accurately observing the effect of different parameters on the slippage of the fluid over the wall of the nanotube, the velocity profile near the wall is shown in a blown-up view. Figures 7(b) and (e) represent the influence of concentration of salt solution on the EOF velocity distribution in soft nanopore. Observing from Figs. 7(a) and (d), it has been concluded that increasing the value of $C_{KCl}$ leads to the decrease of overall electric potential inside the nanopore. Accordingly, the mobility of ionic species, and consequently, the EOF velocity is weakened by increasing $C_{KCl}$. Also, the non-monotonic influence of $\sigma_p$ on velocity profile can be observed from Figs. 7, such that, for both cases of PA and PB grafted nanopores, the initial increase of $\sigma_p$ leads to the increment of velocity plateaus, while further increasing $\sigma_p$ at the interval of [0.2, 0.5] results in the decrease of velocity profile, which is due to the competition between the electroosmotic driving force and hydrodynamic resistance of the PEL against the fluid flow.

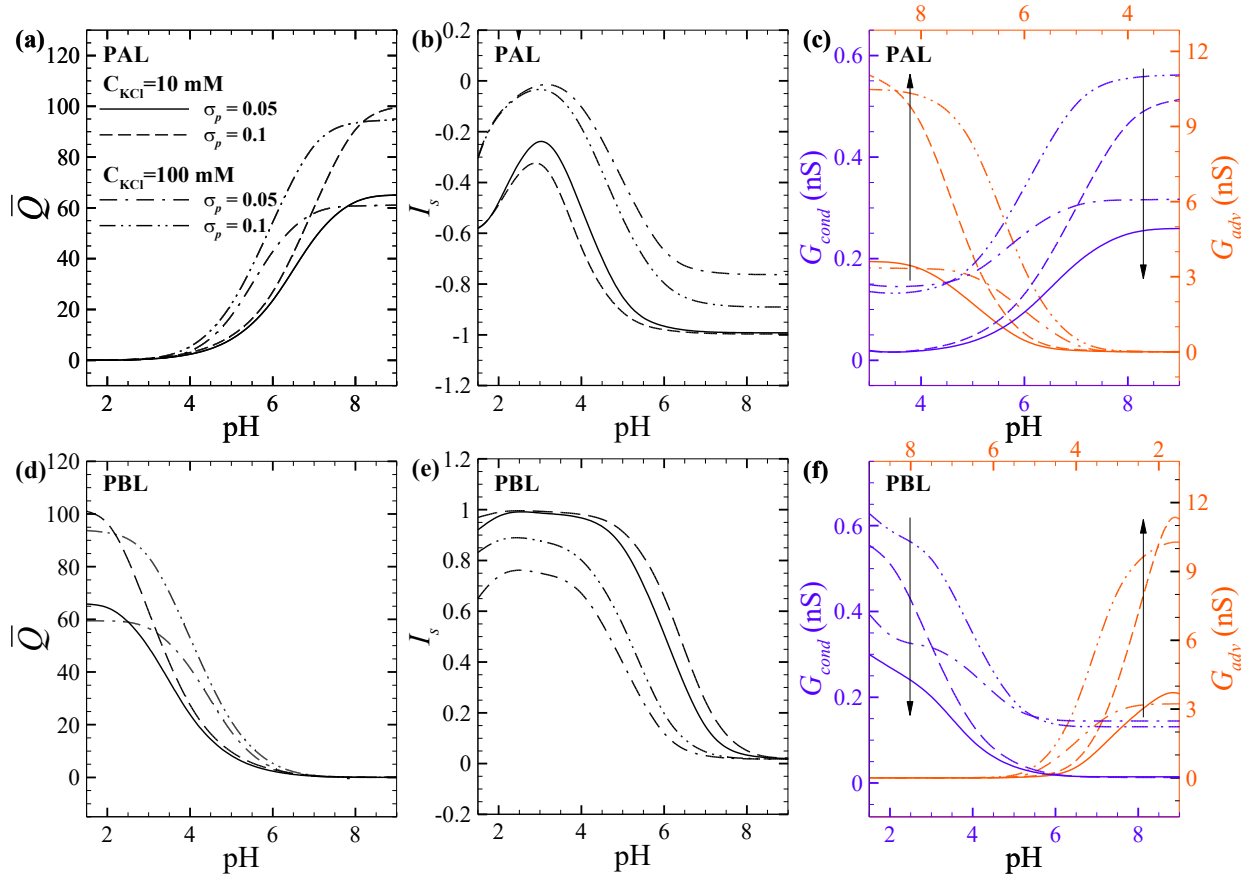

**Fig. 8:** The variation of **(a, d)** volumetric flow rate, **(b, e)** ionic selectivity, and **(c, f)** ionic conduction and advection with pH of the electrolyte in soft nanopore at two values of grafting density of PE-brushes and salt concentration. All panels are in the same parametric conditions, and top/bottom panels correspond to the case of nanopore grafted with PAL/ PBL.

At this stage, we quantitatively investigate the characteristics of EOF and soft nanopore, namely, volumetric flow rate, ionic selectivity, ionic conduction and advection at different values of controlling parameters. As the extreme cases involving large PEL grafting densities are of fewer practical nanofluidic applications [18], we focus here on the cases where $\sigma_p < 0.2$ nm$^{-2}$; however, for higher ranges of $\sigma_p$, our calculations reveal an approximately same trend of characteristics properties of EOF with respect to pH of the electrolyte with various levels. Figures 8(a)/(d) illustrate the dependence of $\bar{Q}$ on pH of the electrolyte at different values of the molar concentration of salt solution and the grafting density of nanopore modified by PAL/PBL. It can be seen that at extreme values of pH of the electrolyte, the $\bar{Q}$ reaches its extremum. Totally, increasing pH promotes the electroosmotic flow rate in nanopores grafted with PAL, while it lowers the amount of $\bar{Q}$ in the PB grafted nanopore. Moreover, as increasing the value of $\sigma_p$ simultaneously increases the number of charge monomer inside the PA and PB grafted nanopore, the strength of electroosmotic body force, and consequently, the volumetric flow rate, increases with increasing $\sigma_p$ which is evident from Figs. 8(a) and (d). However, increasing the concentration of salt solution suppresses the curves of $\bar{Q}$ at large/low values of pH for the case of PAL/PBL grafted nanopore. The reason of such variation can be explained by the acceleration of ionization reaction of polymer brushes in favor of charged monomers in presence of higher concentrations of salt solutions resulting in an increase of $\langle f \rangle$ (see Figs. 4(c, d)) followed by the increase of EOF strength.

The dependence of ionic selectivity of the nanopore on pH of the electrolyte at different values of $\sigma_p$ and $C_{KCl}$ is represented in Figs. 8(b, e). According to the sign of the charges of the ionized monomers within the PAL and PBL, it can be said and also seen from these figures that, the PA grafted nanopore possesses the negative values of $I_s$ (*i.e.*, cation selective nanopore), while the other type of the soft nanopore gets the positive values of $I_s$ (*i.e.*, anion selective nanopore). It means that the majority of the ionic current in the PA grafted nanopore is constituted of positive ions, while for the other case, the negative ions acquire the prevailing concentrations in the net ionic current over the ions of opposite sign. It is noted that the non-monotonic variation of $I_s$ with pH of the electrolyte for both cases of PA and PB grafted nanopore attributes to the dual effect of pH on the nanopore ionic transport. More precisely, the increase of cationic selectivity of the PA grafted nanopore with respect to pH in the range pH $\gtrsim$ 3 in mainly attributed to the increase of the number of negative charged monomers inside the soft nanopore (see Fig. 6(c)). However, in the range pH < 3, further increasing the concentration of $H^+$ results in a broadening of the EDL through the nanopore making an approximately unipolar medium in favor of cations. For the case of PB grafted nanopore, decreasing the ionic selectivity with respect to pH of the electrolyte in the range pH $\gtrsim$ 2 is mainly due to decreasing the degree of ionization of the PBL (see Fig. 6(c)), while for lower values of pH, the monomer of the PBL is mainly charged and anionic selectivity nature of the PB grafted nanopore is deteriorated by the excess concentration of $H^+$ through the nanopore. Moreover, the coupled effects of the bulk concentration and pH of the salt solution on the $G_{cond}$ and $G_{adv}$ is presented in Figs. 8(c, f) in which, the increasing/decreasing dependence of these parameters with pH for PA/PB grafted nanopore, and also, an approximately increasing behavior of $G_{cond}$ and $G_{adv}$ with $C_{KCl}$ and $\sigma_p$ can be observed from these figures.

Then, we calculate the variation of the $\bar{Q}$, $I_s$, $G_{cond}$ and $G_{adv}$ over a wide range of $C_{KCl}$ and different values of pH of the electrolyte, and we show the corresponding results in Fig. 9(a-f). As can be seen, $\bar{Q}$ is an ascending function of $C_{KCl}$ at the intermediate values of pH for EOF in both PA and PB grafted nanopores. However, the rate of variation of $\bar{Q}$ with $C_{KCl}$ increases with increasing/decreasing pH for EOF through PA/PB grafted nanopore. The reasons of such behaviors of $\bar{Q}$ with respect to $C_{KCl}$ and pH are attributed to the change in the electrostatic characteristics of soft nanopore as a results of variation of these parameters which has been presented in the discussion of Figs. 5 and 6. The importance of concentration of salt solution on the level of ionic selectivity of the PA and PB grafted nanopores is shown in Figs. 9(b, e). We can see that by increasing the salt concentration, the cationic/anionic selectivity of PA/PB grafted nanopore decreases, and the situation is worse for lower/higher values of pH of the electrolyte solution.

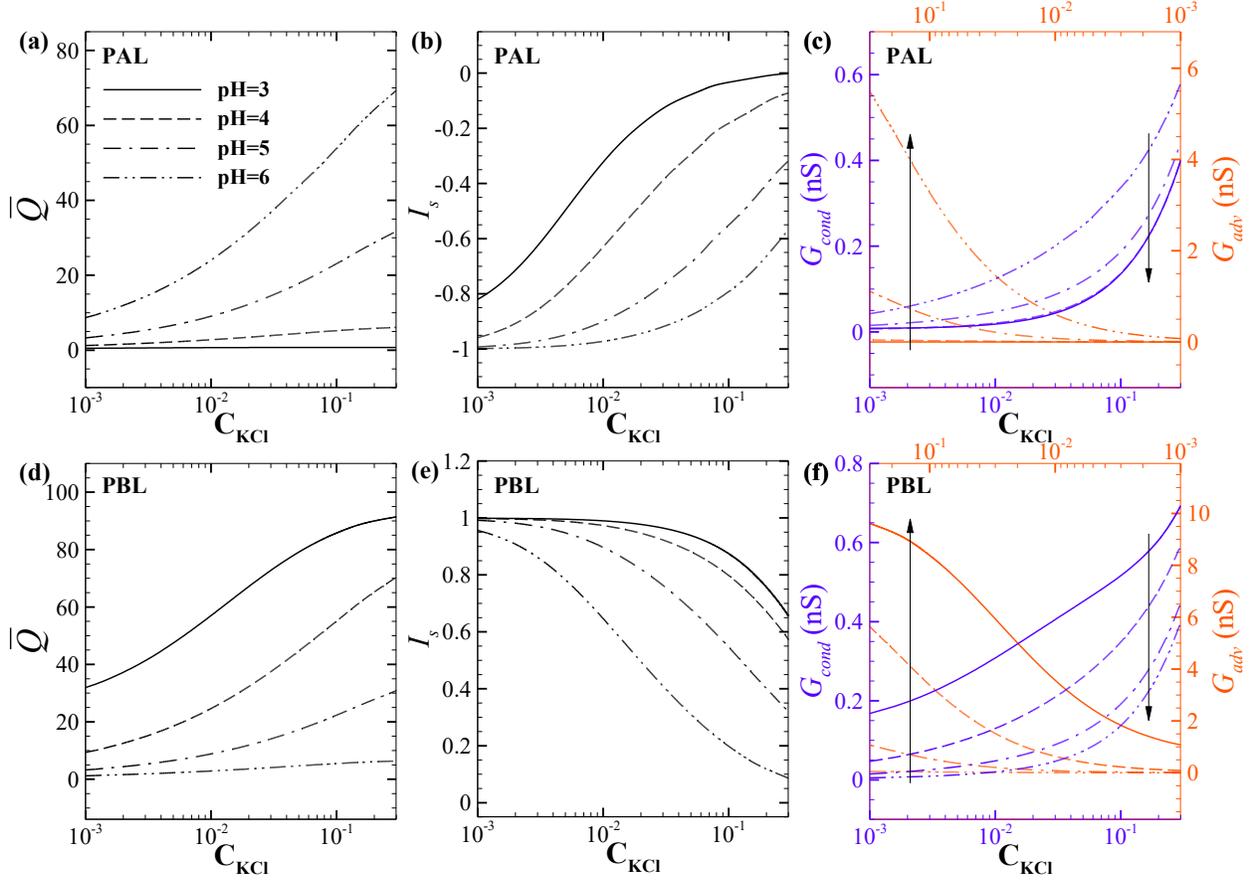

**Fig. 9:** The variation of **(a, d)** volumetric flow rate, **(b, e)** ionic selectivity, and **(c, f)** ionic conduction and advection with molar concentration of the salt solution in soft nanopore at different values of the pH of the electrolyte. All panels are in the same parametric conditions, and top/bottom panels correspond to the case of nanopore grafted with PAL/PBL.

Moreover, we have calculated the variation of ionic conduction and advection by EOF in the PAL and PBL grafted nanopores over a wide range of $C_{KCl}$, and the corresponding results for different values of the pH of the electrolyte are shown in Figs. (c) and (f), respectively. As is evident, both dependent quantities are the ascending function of $C_{KCl}$. However, the rate of variation of the mentioned quantities strongly depends on the pH of the electrolyte, such that, by increasing/decreasing the value of pH, the level and the rate of the growth of the curves of $G_{cond}$ and $G_{adv}$ with respect to $C_{KCl}$ is increased in PA/PB grafted nanopores. The reason of these behaviors are based on this fact that the rate of protonation/deprotonation of two kinds of monomers depends on the local concentrations of the hydronium and hydroxyl ions, which their degree of charge may be strongly affected by the concentration of the salt solution. Finally, we can conclude from the results shown in Figs. 8 and 9 that the volumetric flow rate, ionic conductivity and the selectively of the nanopore can be accurately tuned by properly adjusting the controlling parameters, namely, the grafting density of the PEL ($\sigma_p$), pH of the electrolyte and the concentration of the electrolyte solution.

## 8. Conclusions

Using the molecular theory approach, we generalize our previous analysis of EK flows in soft nanofluidics [44] to include the combined effects of the Born energy arising from the variation of permittivity, fluid-wall slippage, pH of the electrolyte, grafting density of the weak PE brushes, ion partitioning and ionic size effects on the electrokinetic flow characteristics in PE grafted nanopores. In this regard, the Rotational Isomeric State model was employed for the generation of the conformations of the PE brushes, and the generated conformations were used in performing the minimization of the free energy functional of the system. The resulting mobile ionic species, polymer and water distribution functions are evaluated numerically by solving the derived equations from the minimization of the free energy functional of the system, and then, the velocity field was obtained in the process of solving the Navier–Stokes–Brinkman equation. Afterwards, the present results of ionic conductivity were validated by comparing the existing experimental data for a nanopore grafted with 4PVP brushes used as synthetic proton-gated ion channel. Also, the results of ions and potential distribution, degree of ionization of polyelectrolyte (PE) brushes, velocity profile, volumetric flow rate, ionic selectivity, ionic conduction and advection by electroosmotic flow were separately presented and discussed for these for two cases of nanopore grafted with poly-acid and poly-base layers. To the best of our knowledge, it is the first time that the effects of all aforementioned factors on the characteristics of electroosmotic flow through PE grafted nanopores are investigated in the conformation-dependent fashion. In the following, we summarize our key findings obtained from the observation of the presented results:

- *Electrostatics*: the type of the polymer chains determines the concentration of each ionic species in soft nanopore, where for the case of nanopore grafted with PAL, the concentration of cations is significantly larger than that of anions, and vise versa for PB grafted nanopore. For both cases of nanopore grafted with PAL and PBL, by increasing the ionic strength of the electrolyte, the magnitude of the average electric potential inside the soft nanopore is decreased, while the degree of polymer charge is increased. Increasing the pH of the electrolyte results in an increase/decrease of the degree of the charged sites within the PA/PB brushes, Moreover, ignoring the local pH-regulation of PE-molecules leads to the signification inaccuracy of the conductivity results specially at intermediate pH intervals.

- *Velocity field*: increasing the pH of the electrolyte results in the rising of the velocity plateaus through the nanopore grafted with PAL, while the reverse dependence of velocity profile on pH exists for the case of PB grafted nanopore. Also, the EOF velocity is weakened by increasing the bulk salt concentration, while, due to the competition between the electroosmotic driving force and hydrodynamic resistance of the PEL against the fluid flow, there exists a non-monotonic influence of PEL grafting density on velocity profile.

- *Volumetric flow rate*: Increasing pH promotes the dimensionless electroosmotic flow rate ($\bar{Q}$) in nanopores grafted with PAL, while it lowers the amount of $\bar{Q}$ in the PB grafted nanopore. Also, the values of $\bar{Q}$ increases with increasing $\sigma_p$ at low ranges of this parameter, while increasing the bulk concentration of salt solution slightly suppresses the amount of $\bar{Q}$ at large/low values of pH for the case of PAL/PBL grafted nanopore. At the intermediate values of pH, $\bar{Q}$ is an ascending function of $C_{KCl}$ for EOF in both PA and PB grafted nanopores. However, the rate of variation of $\bar{Q}$ with $C_{KCl}$ increases with increasing/decreasing pH of the solution for EOF in PA/PB grafted nanopore.

- *Ionic selectivity*: under the specified conditions of the problem, the PA/PB grafted nanopore is in cation/anion selective state. Also, for both cases of PA and PB grafted nanopore, there exists a the non-monotonic variation of ionic selectivity with pH of the electrolyte, such that, there exists a minimum/maximum point in the variation of magnitude of the ion selectivity of PA/PB grafted nanopore with pH of the solution. Moreover, by increasing the bulk salt concentration, the cationic/anionic selectivity of PA/PB grafted nanopore decreases, and the situation is worse for lower/higher values of pH of the electrolyte solution.

- *Ionic conduction and advection*: both ionic conduction ($G_{cond}$) and advection ($G_{adv}$) are approximately the ascending function of the bulk concentration of the salt solution and low range PEL grafting density. However, the rate of variation of these quantities strongly depends on the pH of the electrolyte, such that, by increasing/decreasing the value of pH, the level and the rate of the growth of the curves of $G_{cond}$ and $G_{adv}$ with respect to $C_{KCl}$ are higher as compared to those of lower pH values in PA/PB grafted nanopores.